\documentclass[preprint,11pt,preprintnumbers,nofootinbib]{revtex4}
\usepackage{graphicx}
\usepackage{dcolumn}
\usepackage{bm}
\usepackage{amsfonts}

\begin{document}
\date{\today}
\title{Preheating after Multi-Field Inflation}

\author{Diana Battefeld}
\email{diana.battefeld(AT)helsinki.fi}
\affiliation{Helsinki Institute of Physics,
        P.O. Box 64, University of Helsinki, FIN-00014 Helsinki, Finland}
\affiliation{Laboratoire APC, Bˆatiment Condorcet, 10, Rue Alice Domon et L´eonie Duquet, 75205 Paris Cedex 13
       }

\begin{abstract}
In this note I study preheating after multi-field inflation to assess the feasibility of parametric resonance. An intuitive argument for the suppression of resonances due to dephasing of fields in generic multi-field models is presented. This effect is absent in effective single field models, rendering them inappropriate for the study of preheating.

\vspace{1pc}
\end{abstract}

\maketitle

\section{Introduction}
To explain how the Universe got the high temperature necessary for Big Bang Nucleosynthesis one needs to understand how the ingredients that drove inflation decay into standard model particles. Our current understanding of preheating is primarily based on single field models, i.e. the old theory of reheating (perturbative, slow decay) or parametric resonance models \cite{Traschen:1990sw,Kofman:1997yn,Bassett:2005xm}. The former can be problematic, since more often than not the decay of the inflaton is not complete; hence, resonance mechanisms are currently preferred. Here a matter field, i.e. another scalar field $\chi$, is coupled to the inflaton field $\varphi$, and $\chi$-particles are produced in short bursts while the inflaton oscillates around the minimum of its potential. Particle production can occur whenever the effective mass $m_{eff}$ of $\chi$ becomes light \cite{Kofman:1997yn}, given that the coupling constant is chosen appropriately. If the expansion of the universe is included, stochastic resonance usually occurs \cite{Kofman:1997yn}.

However, the presence of these resonances in multi-field models is largely unexplored, primarily because setups which employ hundreds of fields were proposed  recently with the advent of more realistic scenarios of inflation within string theory; examples are inflation driven by axions as in $\mathcal N$-flation \cite{Dimopoulos:2005ac}, by tachyons \cite{Majumdar:2003kd} or by multiple M5-branes \cite{Becker:2005sg}. A challenge in many models is to prevent reheating of hidden sectors \cite{Green:2007gs}, but even if the inflatons are primarily coupled to standard model fields, little is known about the presence of resonances in setups with more than a few fields \cite{Bassett:2005xm,Bassett:1997gb}. In this note I provide an intuitive argument for the suppression of resonances in generic models of inflation with many fields, summarizing results obtained in \cite{Battefeld:2008bu}.

\section{Suppression of Resonances}
Consider $\mathcal{N}$ uncoupled, scalar inflaton fields $\varphi_i$ that approach their respective minima after inflation. Generically, these fields oscillate with different frequencies, since there is no a-priory reason why their potentials should be identical. Thus, even though all inflatons start oscillating around the same time, that is in phase, they will quickly dephase. This means that oscillations in $\sum_i \varphi_i^2$ are quickly damped away, leaving a smoothly redshifting term.

Consider now a matter field $\chi$ which is coupled quadratically to all of the inflatons with about the same strength, that is via $g^2 \chi^2 \sum_i \varphi_i^2/2$. As a consequence, its effective mass picks up a term proportional to $g^2 \sum_i \varphi_i^2$. If there were only one field, this contribution would vanish twice per oscillation, leading to two short bursts or particle production at certain wave-numbers when the effective mass is small, given that $g$ is chosen appropriately and the matter fields' bare mass is small. However, in the presence of hundreds of dephased fields, $m_{eff}$ never becomes small enough for parametric resonance to occur; heuristically, one could say that $\chi$ particles are simply too heavy to be produced. 

This intuitive argument was verified numerically in \cite{Battefeld:2008bu}\footnote{This study incorporates the expansion of the Universe but ignores backreaction.} within the framework of $\mathcal{N}$-flation, where $\mathcal{N}\sim \mathcal{O}(10^2)$ fields contribute to preheating. In $\mathcal{N}$-flation the inflatons all oscillate with different frequencies, because their masses are distributed according to the Marcenko-Pastur law; therefore they dephase quickly within the first few oscillations. During these initial oscillations  there is some particle production, but the overall amplitude decreases since this amplification is not strong enough to overcome the dilution due to cosmic expansion. After many oscillations some fields can become in phase again for short intervals, leading again to minor resonances, but these are infrequent and weak. This suppression of resonances is of course entirely absent if an effective single field model is used \cite{Battefeld:2008bu}.  

\section{Consequences}
Given that resonances of the above type are suppressed, how can the Universe become hot? One option is the old theory of reheating with all of its shortcomings, such as potentially incomplete decay of the inflaton with undesirable consequences \cite{Battefeld:2008bu,Braden}. Of course, one could also attempt to incorporate more matter fields, each of which only coupled to a single inflaton field. In this case resonances are still feasible at the price of complicating the scenario considerably. Another option is tachyonic preheating via tri-linear interactions \cite{Dufaux:2006ee}; here, the inflatons contribution to the effective mass is proportional to $\sum_i \varphi_i$ so that resonances still occur frequently \cite{Braden}. However, whether such interactions are present in the models of multi-field inflation proposed in the literature or not  is yet to be seen. 

\section{Outlook}
In the presence of many inflaton fields whose collective behavior is not coherent, parametric resonance in models with quadratic interactions are generically suppressed. As a consequence, more involved methods of reheating, such as tachyonic preheating, have to be considered. Further work is needed to incorporate these methods into concrete scenarios of multi-field inflation. In addition, a more detailed study incorporating backreaction is desirable \cite{preparation}.

\begin{acknowledgments}
I would like to thank my collaborator S.~Kawai, as well as T.~Battefeld for comments. D.~B.~is supported by the EU FP6 Marie Curie Research and Training Network "UniverseNet" (MRTN-CT-2006-035863) and thanks the Institut d'etudes Sciencifiques de Cargese for the amazing hospitality.
\end{acknowledgments}

\end{document}